\documentclass{jetpl} %Latex
\twocolumn
\usepackage[T1,T2A]{fontenc}
\usepackage[utf8]{inputenc}
\usepackage[english]{babel}
\DeclareMathOperator{\Df}{{\mathfrak{ D}}}

\renewcommand{\bf}{\textbf}

\DeclareMathOperator{\bx}{\bf x}
\DeclareMathOperator{\bxq}{\bf x_q}

\DeclareMathOperator{\Qs}{Q\Big(\Phi_s,\frac{\delta}{\delta\Pi_s}\Big)}
\DeclareMathOperator{\Qsm}{Q\Big(\Phi_s,-\frac{\delta}{\delta\Pi_s}\Big)}

\title{Local quench within Keldysh technique}

\rtitle{Local quench}

\sodtitle{Local quench}
\begin{document}
	
	\author{A.\,A.\,Radovskaya$^{*}$, 
		A.\,G.\, Semenov$^{*,**}$}
	
	\sodauthor{Radovskaya, Semenov}
	
	%%% author's address(es)
	\address{$^*$ P. N. Lebedev Physical Institute, Moscow 119991, Russia\\
	$^{**}$ Skolkovo Institute of Science and Technology, Moscow 121205, Russia}
	
	\abstract{The problem of quantum scalar field evolution after an instantaneous local perturbation (quench) is considered. A new approach to descriptions of a quench from an arbitrary initial state is developed in the framework of the Keldysh technique. This approach does not require the procedure of the analytical continuation, which can be ambiguous in some cases. The evolution of the energy density after local quench is calculated for a simple case, and its dependence on the interaction region width and the initial conditions is analysed. }	
	\maketitle
	
	\section{Introduction}
	
Investigations of physical phenomena arising during the quantum evolution of systems with a large number of degrees of freedom are interesting both in themselves and in their application to various areas of modern physics, such as condensed matter physics, cosmology, heavy ion collisions, etc. For example, in ultracold atom experiments, it is possible to change the trap configuration and/or scattering length, which drives the system to a nonequilibrium state. This, in turn, allows one to directly observe the quantum evolution of a many-particle system\cite{Berges}.

The initial state of a quantum system is generally given by the density matrix $\hat\rho(t_0)=\hat\rho_0$. In particular, in a state of thermal equilibrium ${\hat\rho_0\sim e^{-\hat H/T} }$, where $\hat H$ is the Hamiltonian of the system and $T$ is the temperature. After the unitary evolution,  the observable, corresponding to the self-adjoint operator $\hat O$, is measured. The average value of the operator at time $t$ is given by a trace with a density matrix $\langle \hat O\rangle_t={\rm tr}(\hat O\hat\rho(t))$.

The initial density matrix $\hat\rho_0$ is not always known in explicit form. Another way to specify the initial state of the system is to describe the correlations that are present at the initial time moment. In other words, by knowledge of all possible averages $\langle\hat A_1...\hat A_n\rangle_{t_0}={\rm tr}(\hat A_1...\hat A_n\hat\rho_0)$ from the complete set of operators $ \{\hat A_i\}$.
	
Moreover, the initial state of a quantum system can be created with the help of a controlled perturbation of a known equilibrium state. In this case, the initial state is determined by the protocol of its preparation. Suppose the system under consideration was in equilibrium with its environment, and then  some of the parameters suddenly changed, for example, the coupling constant or mass. For a new Hamiltonian, the initial state is nonequilibrium one, and the system starts to evolve to a new equilibrium. This process of sudden changes of the parameters of the entire system is called a global quench \cite{Ruggiero, Calabrese2007, Calabrese2016, SotiriadisElett, SotiriadisPhysRev, Das2015, Das2016, Szasz}. If the system is perturbed in the vicinity of some point $\bxq$ by the action of the operator $\hat Q(\bxq)$, then  such a process is called local quench in the literature\cite{AgeevHep,Horvath,Nozaki,Caputa}. It is interesting that in some cases it is equivalent to a geometric quench, which describes the process of two subsystems merging in field theory \cite{ Calabrese2016, Calabrese1}. After the perturbation, the density matrix of the system has the following form:
	
	$$
	\hat\rho(t_0+0)=\frac{\hat Q(\bxq)\hat\rho_0\hat Q^\dag(\bxq)}{{\rm tr}(\hat Q^\dag(\bxq)\hat Q(\bxq)\hat\rho_0)}.
	$$
	
In particular, an operator of the form ${\hat Q(\bxq)=e^{i\hat V(\bxq)}}$ with self-adjoint $\hat V(\bxq)$ can be interpreted as a result of the influence of the instantaneous perturbation $\delta H (t)=-\delta(t-t_0)\hat V(\bxq)$ of the original Hamiltonian.

In recent works \cite{AgeevHep, Nozaki, Caputa} the local quench of the free scalar field theory was studied. In these works, the authors used the procedure of analytical continuation from imaginary time, which sometimes is not transparent and has a number of restrictions on the initial states.

In this paper, we introduce an approach for studying  a local quench in real time using the Keldysh technique. This approach allows one to investigate a wider class of initial states. Moreover, this approach can be straightforwardly  generalised to the case of interacting theories.

The outline of the paper is as follows: The second section briefly describes the Keldysh technique and the semiclassical approximation within its framework. In the third section, the formulation of the local quench problem is given, and its general solution is obtained within the framework of the approach described above. As an example, a quadratic quench for different initial conditions is considered. In the remaining two sections, the results and
and opportunities for further research are discussed.
	
	%%%%%%%%%%%%%%%%%%%%%%%%%%%%%%%%%%%%%%%%%%%
	\section{Semiclassical approximation within the Keldysh technique }%%%%%%%%%%%%%%%%%%%%%%%%%%%%%%
	%%%%%%%%%%%%%%%%%%%%%%%%%%%%%%%%%%%%%%%%%%%

	It is convenient to study nonequilibrium quantum field systems with the help of the Keldysh technique \cite{Berges,Keldysh, Schwinger,Arseev}. In this approach, the average of the operator is evaluated as a trace with a density matrix. Since the evolution of the density matrix over time is determined by two evolution operators, the doubling of the degrees of freedom occurs in the theory. It can be thought of as the evolution of fields forward $\varphi_F$ and backward in time $\varphi_B$ on the Keldysh contour \cite{LeonidovJetp}.
	
	It is convenient to rotate the field basis $\varphi_F,\varphi_B$
	to the basis of the so-called “classical” $\varphi_{cl}$ and “quantum” $\varphi_q$ fields:
	%$\phi_c\equiv \phi_r$ and $\phi_q \equiv \phi_a $) \cite{berg1,GKSV2020,method}
	\begin{gather}
		\varphi_{cl}(x)=\frac12\left(\varphi_{F}(x)+\varphi_{B}(x)\right),\\ \nonumber
		\hbar\varphi_{q}(x)=\varphi_{F}(x)-\varphi_{B}(x).\nonumber
	\end{gather}
	This substitution is especially convenient for the semiclassical expansion of the theory. In addition, the vertices in this new basis look simpler. Then the average of the operator at time $t$ looks like \cite{Radovskaya}:
	\begin{gather}\label{common}
		\langle  O[\hat\varphi(\bx)] \rangle_{t} = 
		\int\mathfrak D\Pi(\bx)\mathfrak D\Phi(\bx)\ \mathcal W[\Phi(\bx),\Pi(\bx)]\\
		\times \int\limits_{i.c.}\mathcal
		D\varphi_{cl}(t,\bx)\int\mathcal D\varphi_q(t,\bx)  O[\varphi_{cl}(t,\bx)]
		e^{\frac{i}{\hbar}S_K[\varphi_{cl},\varphi_q]}\nonumber .
	\end{gather}
	The action on the Keldysh contour is defined as the difference between the actions on the upper and lower parts of the contour 
	$S_K[\varphi_{F},\varphi_{B}] = S[\varphi_F]-S[\varphi_B]$.
	An integral with the notation $i.c.$ means integration over the fields $\varphi_{cl}$ with the following initial conditions
	$\varphi_{cl}(t_0,\bx)=\Phi(\bx)$, $\partial_t\varphi_{cl}(t_0,\bx)=\Pi(\bx)$. Initial conditions for integration over
	$\varphi_q$ are not set.
	
	The Wigner functional is expressed through the initial density matrix of the system; thereby, it defines the properties of this system at the initial time $t_0$:
	\begin{multline}\label{wigner}
		W[\Phi(\bx),\Pi(\bx)] = \int \Df \beta(\bx)  e^{i\int d^{d-1}\bx \beta(\bx)\Pi(\bx)}\\
		\times\langle  \Phi(\bx)
		+ \frac{\hbar}{2}\beta(\bx)|\hat \rho(t_0)|\Phi(\bx) - \frac{\hbar}{2}\beta(\bx)\rangle.
	\end{multline}
	For the scalar theory: 
	\begin{multline}
		S=
		\frac12\int d^dx\left(\partial_{\mu}\varphi(x)\partial^{\mu}\varphi(x)-m^2\varphi^2(x)-\frac {g}{2}\varphi^4(x)\right).\nonumber
	\end{multline}
	Keldysh action is:
	\begin{gather}
		S_K[\varphi_{cl},\varphi_{q}] = - \hbar\int\limits_{t_0}^{\infty} dt \int d^{d-1}\bx\ \Big(
		\varphi_q A[\varphi_{cl}] + \frac{g\hbar^2}{4}\varphi_{cl}\varphi_q^3\Big),\nonumber
		\\
		A[\varphi_{cl}] = (\partial_{\mu}\partial^{\mu}+m^2)\varphi_{cl} + g\varphi_{cl}^3.\label{SKel}
	\end{gather}
 Here $A[\varphi_{cl}] = 0$ is the equation of motion for a scalar field. It selects fields that belong to the classical trajectories. It is easy to see that the semiclassical expansion can be done by expanding the last term $\frac{g\hbar^2}{4}\varphi_{cl}\varphi_q^3$ in the equation (\ref{SKel}):
	\begin{multline}
		e^{-i\frac{ g\hbar^2}{4}\int\limits_{t_0}^{\infty} dt \int d^{d-1}\bx\varphi_{cl}\varphi_q^3} =\\
		1
		-i\frac{g\hbar^2}{4}\int\limits_{t_0}^{\infty} dt \int d^{d-1}\bx\ \varphi_{cl}\varphi_q^3+\cdots
	\end{multline}
 Leading Order of this expansion is known as Classical Statistical Approximation or Classical method.  Taking into account only the first term of  expansion above, the integrals over fields  $\varphi_q$ и $\varphi_c$ can be done, and the result is ( see, for example, \cite{LeonidovEpjc} ):
	\begin{multline} \label{csa}
		\langle O[\hat\varphi(\bx)]\rangle_{t} =\\ 
		\int \mathfrak{D}\Phi(\bx)  \mathfrak{D} 
		\Pi(\bx) W[\Phi(\bx),\Pi(\bx)] O[\phi_c(t,\bx)],
	\end{multline}
	where
	$\phi_{c}$ is the solution of the classical equation of motion:
	\begin{gather}\label{EoM1}
		\big(\partial_{\mu}\partial^{\mu}+m^2\big)\phi_{c}+g\phi_{c}^3 = 0
	\end{gather}
	with the initial values:
	\begin{gather}
		\phi_{c}(t_0,\bx) = \Phi(\bx), \quad
		\partial_t\phi_{c}(t_0,\bx) = \Pi(\bx).
	\end{gather}
 In other words, in order to find the average of an operator, it is necessary to calculate its value on the classical trajectory and average over all possible initial conditions with the weight given by the Wigner functional. It is convenient to denote such averaging as:
	\begin{multline}\label{ic_dots}
		\int \mathfrak{D}\Phi(\bx)  \mathfrak{D} 
		\Pi(\bx) W[\Phi(\bx),\Pi(\bx)]( \dots) \equiv \langle \dots \rangle_{i.c.},
	\end{multline}	
	so the average\eqref{csa} can be rewritten as:
	\begin{equation}\label{ic}
		\langle O[\hat\varphi(\bx)]\rangle_{t} =	\langle O[\phi_c(t,\bx)] \rangle_{i.c.}.
	\end{equation}
It is easy to see that the semiclassical expansion in the Keldysh technique is constructed using the parameter $\hbar^2 g$, therefore for
of a noninteracting system, the classical approximation gives an exact answer. Since this work study quench in a noninteracting system, it is the formula \eqref{csa},\eqref{ic} that will be used later on in the paper. However, the above discussion shows how interaction can be naturally incorporated into the quench investigation.
	
	%%%%%%%%%%%%%%%%%%%%%%%%%%%%%%%%%%%%%
	\section{Local quench}%%%%%%%%%%%%
	%%%%%%%%%%%%%%%%%%%%%%%%%%%%%%%%%%%%%

Let us consider a local perturbation of the system at space point $\bxq$ at time $t_q$ or, in other words, a local quench with the operator:
\begin{gather}\label{local_quench}
	\hat Q(\bxq ) = e^{-i \frac{\alpha}{\hbar} V(\hat\varphi_s(\bxq))}.
\end{gather}
Here, the function $V(\hat\varphi(\bxq))$ depends only on the value of the field in some vicinity near the space point $\bxq$, such that the field
$\hat\varphi_s(\bxq) = \int d^{d-1}\bx \eta(\bx-\bxq) \hat\varphi(\bx)$ is not defined at one point but is "smeared" in the vicinity of this point. Here $\alpha$ is a dimensional parameter describing the magnitude of the perturbation, and $\eta(\bx-\bxq)$ is a smooth function that is non-zero only in a small vicinity of the point $\bxq$. Since the products of field operators at coinciding points are not well defined, we assume that the "smearing" function\ $\eta(\bx-\bxq)$ always has a finite width.
Thus, the operator $\hat Q(\bxq)$ excites the system in some small vicinity of the point $\bxq$. Note that such a perturbation can be obtained if one adds a time delta-function perturbation ${\delta \hat H(t) = \alpha \delta(t-t_q) V(\hat\varphi_s(\bxq))}$ to the original Hamiltonian. It means that, for example, a quench with ${\alpha V(\hat\varphi_s(\bxq)) = g \hat\varphi_s^4(\bxq)}$ corresponds to the instantaneous appearance of interaction in the system at point $\bxq$.

The density matrix, which is responsible for the evolution of the system after such a perturbation, is given by:
\begin{gather}
	\hat\rho(t_0) \to \hat\rho_Q(t_q,\bxq) = \hat Q (\bxq) \hat\rho(t_q) \hat Q^\dag(\bxq)
\end{gather}
Below, we assume that the quench occurred at the initial time equal to zero, $ t_q = t_0 = 0$.
Then, according to the formula \eqref{wigner}, the Wigner functional after a local quench is:
\begin{gather}
	W_Q[\Phi(\bx),\Pi(\bx)] = \int \Df \beta(\bx) e^{i\int d^{d-1}\bx \beta(\bx)\Pi (\bx)}\\
	\times\langle \Phi(\bx)
	+ \frac{\hbar}{2}\beta(\bx)| \hat Q (\bxq) \hat\rho(t_0) \hat Q^\dag(\bxq)|\Phi(\bx) - \frac{\hbar}{2}\beta(\bx)\rangle.
\end{gather}
Note, that
\begin{gather}
	-i\frac{\delta}{\delta \Pi(\bf y)} e^{i\int d^{d-1}\bx \beta(\bx)\Pi(\bx)} = \beta (\bf y) e^{i\int d^{d-1}\bx \beta(\bx)\Pi(\bx)}.
\end{gather}
Then the Wigner functional after quench can be rewritten as:
\begin{multline}
	W_Q[\Phi(\bx),\Pi(\bx)] =
	Q\Big(\Phi_s,\frac{\delta}{\delta\Pi_s}\Big)W[\Phi(\bx),\Pi(\bx)],
\end{multline}
where we introduce the notation:
\begin{gather}\label{QPhiPi}
	\Qs = e^{-i\frac{\alpha}{\hbar}\big( V\big( \Phi_s -i\frac{\hbar}{2}\frac{\delta }{\delta \Pi_s} \big) -
		V\big( \Phi_s + i\frac{\hbar}{2}\frac{\delta}{\delta \Pi_s} \big)
		\big)},\\
	\Phi_s = \int d^{d-1} \bx \eta(\bx-\bxq) \Phi(\bx),\\
	\frac{\delta}{\delta \Pi_s} = \int d^{d-1} \bx \eta(\bx-\bxq)\frac{\delta}{\delta \Pi(\bx)}.
\end{gather}
In these expressions, the dependence on the quench point $\bxq$ enters implicitly through the “smearing” function $\eta(\bx-\bxq)$.

Then, after functional integration by parts, the average of the operator can be written as:
\begin{multline}\label{qcsa}
	\langle \hat O\rangle_{t}^Q =
	\int \mathfrak{D}\Phi(\bx) \mathfrak{D}
	\Pi(\bx) W[\Phi(\bx),\Pi(\bx)]\times\\
	\Qsm O[\phi_c(t,\Phi(\bx),\Pi(\bx))].
\end{multline}
From the above equation, one can see that in order to find the average of the operator after quench, it is necessary to perform summation over the initial conditions with the original Wigner functional, but for a modified observable. Since polynomial observables, such as energy density or correlation functions of fields, are most often considered, in the absence of interaction, only a finite number of variational derivatives $\frac{\delta}{\delta \Pi_s}$ remains nonzero. It means that the developed method gives an exact answer even for nonquadratic perturbations $\hat V \hat \varphi_s(\bxq)$.

The formula \eqref{qcsa} provides the main result of this article. We demonstrate its application below.
	
	%%%%%%%%%%%%%%%%%%%%%%%%%%%%%%%%%%%%%%%%%%%%%%%%%%%%%%%%%%%%%%%%%%
	\subsection{Quench $V(\hat\varphi)= \hat\varphi^2$}%%%%%%%%%%%%
	%%%%%%%%%%%%%%%%%%%%%%%%%%%%%%%%%%%%%%%%%%%%%%%%%%%%%%%%%%%%%%%%%%
	
	Let us perform calculations for  1+1D field theory for simplicity. Consider the local perturbation induced by the operator:
	\begin{equation}\label{Qm2}
		\hat Q (x_q) = e^{-i \frac{\alpha}{\hbar} \hat\varphi_s^2(x_q)}.
	\end{equation}
	If, due to this quench, the mass of the field has changed from $m$ to $M$ in the small vicinity $ \Delta x$ of the point $x_q$ for a short period of time $\Delta t$ , than
	$\alpha \approx\frac{M^2-m^2}{2}\Delta t \Delta x $.

	According to formula \eqref{QPhiPi}:
	\begin{equation}\label{Qphi2}
		\Qsm = e^{2\alpha\Phi_s\cdot\frac{\delta}{\delta\Pi_s}}.
	\end{equation}	
As an observable, let us consider the energy density of the system:
	\begin{equation}
		\varepsilon(t,x) = \frac12 (\partial_t\varphi)^2 + \frac12 (\partial_x\varphi)^2 +\frac12 m^2\varphi^2.
	\end{equation}
The last ingredient that is required in order to use the formula \eqref{qcsa} for evaluation of the average energy density after the action of a local quench is the solution of the classical equation of motion \eqref{EoM1}. It is easy to show that, in the absence of interaction, it is equal to:
	\begin{multline}\label{phi_c}
		\phi_c(t, x) =\\ -\int dy \Big(
		\partial_t G_R(t, x-  y)\Phi( y) +  G_R(t, x-  y )\Pi( y) 	\Big),
	\end{multline}	
where retarded Green function is defined from the retarded solution of equation:
	\begin{equation}
		(\partial_t^2 -\partial_x^2 +m^2)G_R(t,x-x') = -\delta(t)\delta(x-x'),
	\end{equation}	
which is:
	\begin{multline}
		G_R(t,x-x') = 
		-\theta(t) \int  \frac{dp}{2\pi} \frac{\sin(\omega_pt)}{\omega_p}
		e^{-ip( x-x')},\\
		\omega_p = \sqrt{p^2+m^2}.
	\end{multline}	
	
Now one can calculate the average energy density after the quench. For simplicity, we write down the expression for $\phi_c^2(t,x)$. Terms with the derivatives are calculated similarly. Since the classical solution is a linear functional of the initial conditions, it is sufficient to expand the exponent in the expression \eqref{Qphi2} to the second order. All higher variational derivatives are zero.
Therefore, with the help of the formulae \eqref{Qphi2},\eqref{phi_c}, one obtains:
	\begin{multline}\label{Q_phi_2}
		\Qsm\phi_c^2(t,x) = \phi_c^2(t,x) \\-4\alpha\phi_c(t,x)\Phi_s\cdot G_R^s(t,x) + 4\alpha^2\Phi_s^2\cdot \Big( 
		G_R^s(t,x) \Big)^2.
	\end{multline}
Here we introduce the notation for the "smeared"\ retarded Green function:
	\begin{equation}
		G_R^s(t,x) = \int dy\ \eta(y-x_q)G_R(t,x-y).
	\end{equation}
	As a next step, it is necessary to perform averaging over the initial conditions with the Wigner functional. The functional is normalised to unity, the retarded Green function does not depend on the initial conditions, and the average of the classical solutions $\phi_c$ is equal to the Keldysh Green function \cite{Radovskaya}:
	\begin{multline}
		iG_K(t-t',x-x') =\\
		\int \Df \Phi(x) \Df \Pi(x) W[\Phi(x),\Pi(x)] \phi_c(t,x)\phi_c(t',x'),
	\end{multline}
which in a standard way, is determined through the trace with the initial density matrix as:
	\begin{gather}
		iG_K(t-t',x-x') =\frac12 tr \big(\hat \rho(t_0) \{\hat\varphi(t,x),\hat\varphi(t',x') \} \big).\nonumber
	\end{gather}
	
	Using notation \eqref{ic_dots} for the integration over initial conditions, let us define the "smeared"\ Keldysh Green function  $G_K^s(t,x)$ and the constant $\langle\Phi_s^2\rangle_{i.c}$:
	\begin{multline}
		\langle\phi_c(t,x) \Phi_s\rangle_{i.c} =i G_K^s(t,x) \equiv \\ \int dy\ \eta(y-x_q)iG_K(t,x-y),\\
		\langle\Phi_s^2\rangle_{i.c}\equiv \int dydz\ \eta(y-x_q) \eta(z-x_q)iG_K(0,y-z).
		\label{GKs}
	\end{multline}
	Than, for the energy density after quench \eqref{Qm2}, we obtain:
	\begin{multline}\label{energy_density}
		\langle  \hat\varepsilon\rangle_{t}^Q =  \langle  \hat\varepsilon\rangle_{t} -
		2 i \alpha \Big(m^2 G_K^s(t,x)G_R^s(t,x)\\ +\partial_t G_K^s(t,x)\partial_t G_R^s(t,x)
		+ 
		\partial_x G_K^s(t,x)\partial_x G_R^s(t,x) \Big)\\
		+	2  \alpha^2 \langle\Phi_s^2\rangle_{i.c}\Big(m^2\big( G_R^s(t,x) \big)^2
		+ \big(\partial_t G_R^s(t,x) \big)^2\\ + \big(\partial_x G_R^s(t,x) \big)^2 \Big).
	\end{multline}

Energy density \eqref{energy_density} is a real value.
The imaginary unity is included in the definition of the Keldysh Green function \eqref{G_K}.
The Keldysh Green function is singular at coinciding points. However, the constant $\langle\Phi_s^2\rangle_{i.c}$ is regularised with the help of the "smearing"\ function $\eta(x-x_q)$. This function was introduced in the quench definition \eqref{local_quench} exactly to eliminate such a divergence. Its physical meaning is that the energy is released not exactly at the point $x_q$, but in a certain vicinity specified by the “smearing” function. Therefore, the final answer depends on this function and diverges if it approaches the delta-function.

The explicit form of the Keldysh Green function depends on the initial conditions of the problem, and if they are described by a single-particle distribution function $f_p$, for the free theory, it is equal to:
	\begin{multline}\label{G_K}
		iG_K(t,x-x') =\hbar \int  \frac{dp}{2\pi} \frac{\cos(\omega_pt)}{2\omega_p}(2f_p+1)
		e^{-ip( x-x')}.
	\end{multline}
	With the help of the explicit form of the Green functions, it is easy to calculate the total energy that the system received after the quench:
	\begin{equation}\label{deltaE}
		\delta E = \int dx \big(	\langle  \hat\varepsilon\rangle_{t}^Q -	\langle  \hat\varepsilon\rangle_{t}\big) =
		2\alpha^2 \langle\Phi_s^2\rangle_{i.c}\int dy \eta^2(y).
	\end{equation}
	
Note that the expression \eqref{energy_density} contains three terms. The first one represents the energy density of the system before the quench. This term can be divergent, as it happens with vacuum energy in quantum field theory. This divergence is not related to the problem under consideration and can be eliminated with the help of standard methods of quantum field theory \cite{Bogolubov}. The second term (proportional to $\alpha$) corresponds to the linear response of the system to a local disturbance. As it follows from the Kubo formula, this term describes the redistribution of energy between different parts of the system and does not contribute to the total energy absorbed by the system \eqref{deltaE}. All the energy absorbed by the system after quench is described by the third term of the expression \eqref{Qm2}.

The total energy \eqref{deltaE} depends significantly on the "smearing"\ function $\eta(x-x_q)$. In the case of a local quench, this function is nonzero only in a small vicinity of the point $x_q$ with size equal to $\epsilon$ and tends to $\delta(x-x_q)$ at $\epsilon \to 0$. The "smearing"\   function is included in the expression for
$\langle\Phi_s^2\rangle_{i.c}$. The Keldysh Green function in coinciding points diverges in a way standard for quantum field theory and requires regularisation. Therefore, at $\epsilon = 0$, the mean square of the field $\langle\Phi_s^2\rangle_{i.c}$ depends on the ultraviolet scale $\Lambda$. In 1+1D scalar field theory, this dependence is logarithmic one $ \langle\Phi_s^2\rangle_{i.c} \sim \log \frac{\Lambda}{m}$. If $\epsilon$ is finite, then
$ \langle\Phi_s^2\rangle_{i.c}$ converges, and in the case of $\epsilon \Lambda \gg 1 $, the ultraviolet behaviour of the theory becomes insufficient.
Moreover, the total energy \eqref{deltaE} absorbed by the system contains an explicit integral of $\eta^2(x)$. Simple dimensional estimates (or an explicit calculation with a Gaussian "smearing"\   function) give $\int dx\ \eta^2(x) \sim \frac{1}{\epsilon}$. Therefore, the most singular contribution to the total energy is:
	\begin{equation}
		\delta E \sim \frac{1}{\epsilon} \log\left( \frac{\min\left(\Lambda, \epsilon^{-1}\right)}{m}\right).
		\label{Eepsolon}
	\end{equation}

Note that the approach used in this work can be applied to any initial state of the system. Figure 1 (a,c) shows the energy density of the system after quench \eqref{energy_density} for the vacuum initial state ($f_p =0$), and Figure 1 (b,d) - for the thermal initial state when the system is characterised by the temperature $T\gtrsim m$ and the Bose distribution function $f_p = (e^{\hbar \omega_p /T}-1)^{-1}$ . It can be seen that the higher the temperature, the more excited the medium is under the action of a local disturbance. This effect is a direct manifestation of the Bose statistics for the problem under consideration. Local quench for systems that are prepared in a termal initial state was also considered in \cite{Caputa}.

	\begin{figure}
		\includegraphics[width = 1\columnwidth]{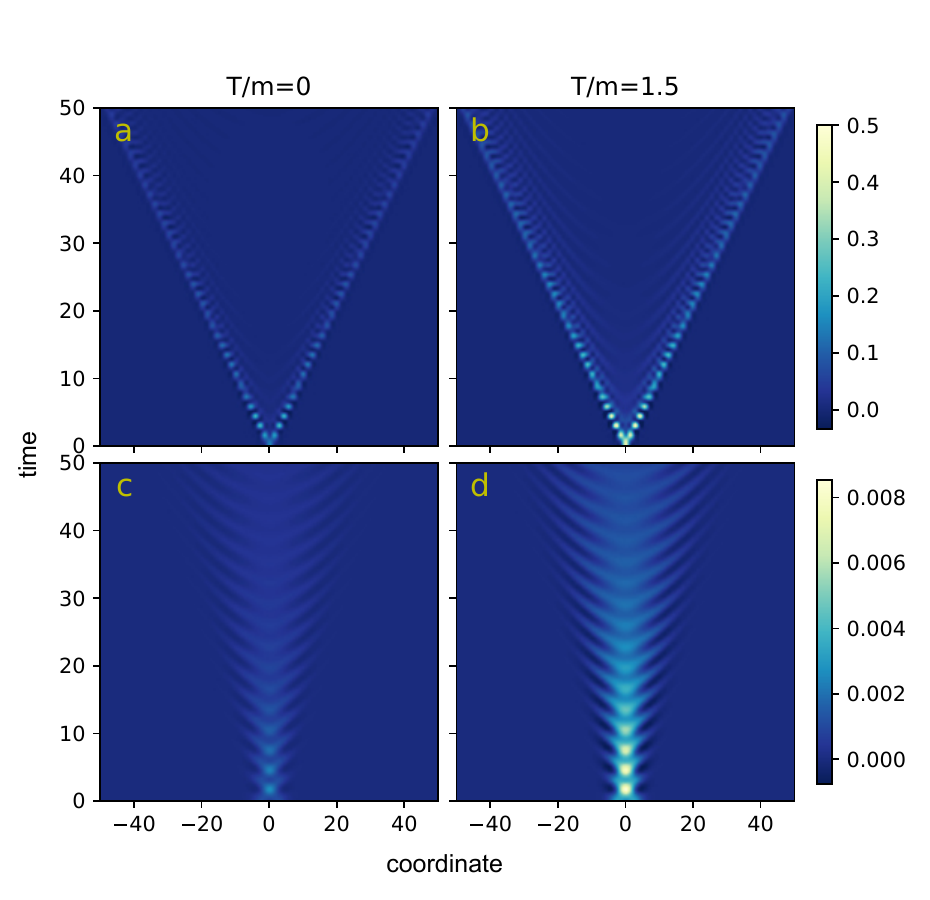}
		\caption{Fig. 1: Distribution of energy density in space depending on time after a local quench at the point $x_q=0$. The thermal state with temperatures $T = 0$ (a,c) and $T = 1.5 m$ (b,d) was chosen as the initial state of the system. The "smearing"\ function is assumed to be Gaussian with a width of $\epsilon m = 0.25 $ (a,b) and $\epsilon m =2 $ (c,d). Position and time are measured in units of inverse mass. }
		\label{fig1}
	\end{figure}
	
The upper and lower pictures in Fig. 1 differ in the value of the width of the "smearing"\ function $\epsilon$. This value determines the maximum momentum of particles created during quench as $p_{\rm max}\sim \frac{1}{\epsilon}$ . This allows us to estimate the typical speed of propagation of the energy density disturbance front as the maximum group velocity of particles $v_{\rm max}$:
	\begin{gather}
		v_{\rm max} = \frac{\partial \omega_p}{\partial p} \sim \frac{p_{\rm max}}{\sqrt{p_{\rm max}^2+m^2}}\sim 
		\frac{1}{\sqrt{1+m^2\epsilon^2}}
	\end{gather}
	
If the area of disturbance is small ($\epsilon \ll m^{-1}$), then such a disturbance propagates with the speed of light ${v_{\rm max} \sim 1}$ (Fig. 1 (a,b)). In the opposite case ($\epsilon \gg m^{-1}$), the group velocity is small $v_{\rm max} \sim \frac{1}{m\epsilon}$ and the front propagates more slowly (Fig. 1 (c ,d)). A similar regime change was observed in the work of \cite{AgeevHep} for the quench $\hat Q \sim \hat \varphi$. It seems that this behaviour is universal and does not depend on the initial state or the specific type of quench operator.

	%%%%%%%%%%%%%%%%%%%%%%%%%%%%%%%%%%%%%%%%%%%%%%%%%%%%%%%%%%%%%%%%%%%%%%%%%%%%%%%%%
	\section{Discussion}%%%%%%%%%%%%%%%%%%%%%%%%%%
	%%%%%%%%%%%%%%%%%%%%%%%%%%%%%%%%%%%%%%%%%%%%%%%%%%%%%%%%%%%%%%%%%%%%%%%%%%%%%%%%%

In the previous section, we discuss the evolution of energy density after the quench of the form $\hat Q (x_q) = e^{-i \frac{\alpha}{\hbar} \hat\varphi_s^2(x_q)}$. It can be shown that for a more general form of the quench operator ${ \hat Q(x_q ) = e^{-i \frac{\alpha}{\hbar} V(\hat\varphi_s(x_q))} }$ with arbitrary $V( \hat\varphi_s(x_q))$, the energy density after quench is given by the expression:
	\begin{multline}\label{general_e}
		\langle  \hat\varepsilon\rangle_{t}^Q =  \langle  \hat\varepsilon\rangle_{t} -
		\alpha \Big(m^2 G_R^s(t,x) \big \langle \phi_c(t,x) V'(\Phi_s) \big\rangle_{i.c.}\\
		+\partial_t G_R^s(t,x)\big\langle \partial_t\phi_c(t,x) V'(\Phi_s) \big\rangle_{i.c.}\\
		+ 
		\partial_x G_R^s(t,x) \big\langle \partial_x\phi_c(t,x) V'(\Phi_s)\big \rangle_{i.c.}\Big)\\
		+	
		\frac12  \alpha^2 \big\langle V'(\Phi_s) V'(\Phi_s)\big\rangle_{i.c}\Big(m^2\big( G_R^s(t,x) \big)^2\\
		+ \big(\partial_t G_R^s(t,x) \big)^2 + \big(\partial_x G_R^s(t,x) \big)^2 \Big).
	\end{multline}	
The averages over the initial conditions $\langle\dots \rangle_{i.c}$ entering in this expression can be given not only by pair correlators as in the example considered above. In this case, the typical behaviour of the energy density directly depends on the correlations presented in the system at the initial time moment.

The expression \eqref{general_e} allows us to compare our results with those obtained  earlier  in conformal field theory.
Within the framework of conformal theory, the case when $\hat Q(x)$ is a primary field can be effectively analysed. As an example, we consider the vertex operator ${\hat Q(x)=\hat {\mathcal V}_\alpha(x)=:e^{i\alpha\hat\varphi(x)}:}$ with conformal dimensions $h =\bar h =\alpha^2/(8\pi)$ \cite{Mussardo}. Here $:...:$ denotes normal ordering. In the framework of the Keldysh technique, we should consider the case of the potential $V(\varphi)=-\varphi$ and the vacuum initial state $T=0$ ($f_p=0$).
Moreover, it is necessary to choose the "smearing"\ function in the form:
	\begin{equation}
		\eta(x)=\int\frac{dp}{2\pi} e^{ipx-\epsilon\omega_p}=\frac{m\epsilon}{\pi\sqrt{x^2+\epsilon^2}}K_1\left(m\sqrt{x^2+\epsilon^2}\right),
		\label{smeared}
	\end{equation}
	where $K_\nu(z)$ is the MacDonald function, and $\epsilon$ -- is the small parameter  (the width of the "smearing"\ function). 
The proof of this statement goes beyond the letter format, so we have included it in the supplementary materials (see supplementary material at). Moreover, in the supplementary materials, we consider a quench operator of the form $\hat Q(x)=\hat\varphi(x)$, which is not the primary one; however, it was discussed in detail in the work \cite{AgeevHep}.

For a potential of the form $V(\varphi)=-\varphi$ the expression \eqref{general_e} reduces to the following:
	\begin{multline}
		\langle  \hat\varepsilon\rangle_{t}^Q =  \langle  \hat\varepsilon\rangle_{t}+\frac12  \alpha^2\Big(m^2\big( G_R^s(t,x) \big)^2 \\+ \big(\partial_t G_R^s(t,x) \big)^2 
		+ \big(\partial_x G_R^s(t,x) \big)^2 \Big).
	\end{multline}	
	For the "smearing"\ function  \eqref{smeared} all integrals can be done analytically. In particular, the "smeared"\ retarded Green function has the form: 
	\begin{multline}
		G_R^s(t,x) = \int dy\ \eta(y-x_q)G_R(t,x-y)  \\
		=-\theta(t) \int  \frac{dp}{2\pi} \frac{\sin(\omega_pt)}{\omega_p}
		e^{-ip(x-x_q)-\epsilon\omega_p}\\
		=\frac{i}{2\pi}\theta(t)\left(K_0\left(m\sqrt{(x-x_q)^2+(\epsilon-it)^2}\right)\right. \\
		\left. -K_0\left(m\sqrt{(x-x_q)^2+(\epsilon+it)^2}\right)\right).
		\label{GRs1}
	\end{multline}
The answer looks especially simple in the massless case  $m\to 0$ ( which is described by the conformal field theory with the central charge  $c=1$), then:
	\begin{equation}
		G_R^s(t,x) =\frac{i}{4\pi}\log\left(\frac{(x-x_q)^2+(\epsilon+it)^2}{(x-x_q)^2+(\epsilon-it)^2}\right),
		\label{GRs2}
	\end{equation}
and the energy density after the local quench is:
	\begin{multline}
		\langle  \hat\varepsilon\rangle_{t}^Q =  \langle  \hat\varepsilon\rangle_{t}+\frac{\alpha^2}{4\pi^2}\left(\frac{\epsilon^2}{((x-x_q-t)^2+\epsilon^2)^2}\right. \\ \left.+\frac{\epsilon^2}{((x-x_q+t)^2+\epsilon^2)^2}\right).\label{result}
	\end{multline}
This answer completely coincides with the results of the work of \cite{Caputa2} (formula (34)) and the work of \cite{AgeevHep} (formulae (2.7,2.9)) taking into account the choice of field normalisation and definition of the energy density.

Next, consider the problem of divergences that can arise during the calculation of the energy density after a local quench of the form $ \hat Q(x_q ) = e^{-i \frac{\alpha}{\hbar} V(\hat\varphi_s(x_q)) }$. For a finite width of the "smearing"\ function $\epsilon$, the "smeared" retarded Green functions $G^s_R(t,x)$ entering in the expression \eqref{general_e} do not contain singularities on the real axis (see, for example, the expressions \eqref {GRs1},\eqref{GRs2}), therefore, all possible divergences can only arise during the calculation of the averages over the initial conditions. In the general case, these averages are defined by the form of the Wigner functional, which requires a special, detailed investigation. However, for Gaussian initial conditions, analysis of the divergences can be performed with the help of Wick's theorem. In this case, two types of pairings arise: $\langle \Phi_s^2 \rangle_{i.c.}$ and
$G_K^s = \langle \phi_{c} \Phi_s\rangle_{i.c.}$. It can be shown that for a finite $\epsilon$, the integrals entering the expressions \eqref{GKs} are also convergent. Therefore, for Gaussian initial conditions, the final result for the energy density after the quench under consideration is finite, at least for the polynomial $V(\hat\varphi_s)$. Note that  it is important to distinguish between the width of the "smearing"\ function $\epsilon$ and the ultraviolet cutoff parameter of the theory $\Lambda$ \eqref{Eepsolon}. If $\epsilon \gg \Lambda^{-1}$, the degrees of freedom with energy $\sim \Lambda$ are not excited after quench, which means the final result is not sensitive to the ultraviolet behaviour of the theory.

	%%%%%%%%%%%%%%%%%%%%%%%%%%%%%%%%%%%%%%%%%%%%%
	\section{Conclusion}%%%%%%%%%%%%%%%%%%%%%%%%%%
	%%%%%%%%%%%%%%%%%%%%%%%%%%%%%%%%%%%%%%%%%%%%%%
In this work, we propose a new approach for the description of a local perturbation (quench) in scalar field theory with the help of the Keldysh technique. This approach does not use the analytical continuation procedure, which in some cases may be ambiguous \cite{AgeevHep}.
Moreover, the method presented in the work allows us to consider systems with an arbitrary initial state.

For the quench $\hat Q (x_q) = e^{-i \frac{\alpha}{\hbar} \hat\varphi_s^2(x_q)}$, the evolution of the energy density was calculated for both the vacuum initial state and the state with an arbitrary initial distribution function $f_p$. Two regimes of propagation of the disturbance front are described, depending on the size of the local disturbance region (the width of the “smearing”\ function $\epsilon$).

The approach to the description of the dynamics of a system after an instantaneous local perturbation obtained in this work can be generalised to the case of nonzero interaction, at least for the semiclassical approximation within the Keldysh technique. This is a topic for further investigation.

\newpage
\onecolumn

	\title{Supplementary Material to the article  "Local quench within Keldysh technique"}
	\maketitle
	\section{Introduction}
	
	 The aim of this supplementary material is to demonstrate that the new formalism developed in the main article is in agreement with the previous calculations performed in the framework of the conformal field theory. Let us consider the example of a free scalar field, which can be described by the conformal field theory with the central charge $c=1$. In order to perform these calculations, we use field mode expansion (mass $m$ is finite for the intermediate calculations, but it is set to zero at the end).
	\begin{equation}
		\hat\varphi(x)=\int\frac{dp}{2\pi}\frac{1}{\sqrt{2\omega_{p}}}\left(\hat a_{p} e^{ipx}+\hat a_{p}^\dag e^{-ipx}\right)\equiv\hat\phi(x)+\hat\phi^\dag(x).
	\end{equation}
Here, we define positive and negative frequency parts. The former contains only field annihilation operators; the later - only creation operators. Creation and annihilation operators have the following canonical commutation relations: $[\hat a_p,\hat a_q^\dag]=2\pi\delta(p-q)$. As an initial condition before a quench we consider the vacuum state $|0\rangle$, which is destroyed by any annihilation operator $\hat a_p|0\rangle=0$, which means that $\hat\phi(x)|0\rangle=0$. After a local quench the state became (see, for example \cite{AgeevHep-s,Caputa-s,Calabrese2016-s,Caputa2-s,Nozaki-s}):
	\begin{equation}
		|\psi_0\rangle=\mathcal N e^{-\epsilon\hat H}\hat Q(x_q)|0\rangle.
	\end{equation}
In the above expression, a small $\epsilon$ acts as the UV(ultraviolet) regularisator, $\mathcal N$ -- normalisation constant, such as $\langle\psi_0|\psi_0\rangle=1$.

 The Hamiltonian of a free scalar theory is:
	\begin{equation}
		\hat H=\int\frac{dp}{2\pi}\omega_p\hat a_p^\dag\hat a_p.
	\end{equation} 

In the framework of the conformal field theory, it is more convenient if $\hat Q(x)$ is a primary operator. As an example of such an operator, we consider the vertex operator $\hat Q(x)=\hat {\mathcal V}_\alpha(x)=:e^{i\alpha\hat\varphi(x)}:$, where with $:...:$ the normal ordering is denoted. Moreover, as a second example, we consider the quench operator $\hat Q(x)=\hat\varphi(x)$, which is not a primary operator; however, it was thoroughly investigated in the paper\cite{AgeevHep-s}.

	\section{Vertex operator quench}

Let us consider a state that is created after a local quench by the vertex operator $\hat {\mathcal V}_\alpha(x)$:
	\begin{multline}
		|\psi_0\rangle=\mathcal N e^{-\epsilon\hat H}:e^{i\alpha\hat\varphi(x_q)}:|0\rangle=\mathcal N e^{-\epsilon\hat H}e^{i\alpha\hat\phi^\dag(x_q)}e^{i\alpha\hat\phi(x_q)}|0\rangle=\mathcal N e^{-\epsilon\hat H}e^{i\alpha\hat\phi^\dag(x_q)}|0\rangle\\=\mathcal N e^{-\epsilon\hat H}e^{i\alpha\hat\phi^\dag(x_q)}e^{\epsilon\hat H}|0\rangle=\mathcal N e^{i\alpha e^{-\epsilon\hat H}\hat\phi^\dag(x_q)e^{\epsilon\hat H}}|0\rangle.	\label{state}
	\end{multline}
	With the help of commutation relation $e^{-\epsilon\hat H}\hat a_p^\dag e^{\epsilon\hat H}=\hat a_p^\dag e^{-\epsilon\omega_p}$ one can show that
	\begin{equation}
		e^{-\epsilon\hat H}\hat\phi^\dag(x_q)e^{\epsilon\hat H}=\int\frac{dp}{2\pi}\frac{1}{\sqrt{2\omega_{p}}}\hat a_{p}^\dag e^{-ipx_q-\epsilon\omega_p}.
	\end{equation}
	Let us introduce the "smearing"\ function:
	\begin{equation}
		\eta(x)=\int\frac{dp}{2\pi} e^{ipx-\epsilon\omega_p}=\frac{m\epsilon}{\pi\sqrt{x^2+\epsilon^2}}K_1\left(m\sqrt{x^2+\epsilon^2}\right),
		\label{smeared-s}
	\end{equation}
 here $K_\nu(z)$ is the Macdonald function.  With the help of this function, previous relation can be rewritten as:
	\begin{equation}
		e^{-\epsilon\hat H}\hat\phi^\dag(x_q)e^{\epsilon\hat H}=\int dx\ \eta(x-x_q)\phi^\dag(x)\equiv \hat \phi_s^\dag(x_q).
	\end{equation}
	In other words, for the problem under consideration, the action of the regularisation multiplier $e^{-\epsilon\hat H}$ can be mimicked by the introduction of a certain "smearing"\ function (\ref{smeared-s}). Than for the equation (\ref{state}) we have
	\begin{equation}
		|\psi_0\rangle=\mathcal N e^{i\alpha e^{-\epsilon\hat H}\hat\phi^\dag(x_q)e^{\epsilon\hat H}}|0\rangle=\mathcal N e^{i\alpha \hat \phi_s^\dag(x_q) }|0\rangle = \mathcal N e^{i\alpha \hat \phi_s^\dag(x_q) }e^{i\alpha \hat \phi_s(x_q) }|0\rangle = e^{i\alpha (\hat \phi_s^\dag(x_q)+\hat \phi_s(x_q) )}|0\rangle = e^{i\alpha \hat \varphi_s(x_q)}|0\rangle.
	\end{equation}
In the above equation, we use the fact that the commutator of $\hat\phi_s(x)$ and $\hat\phi_s^\dag(x)$ is a number, which can be combined with the normalisation constant $\mathcal N$. Therefore, we have shown that the problem of the quench induced by the vertex operator can be considered as a particular case of the problem solved in the main article. Namely, it is the case of the potential $V(\varphi)=-\varphi$ and vacuum initial state $T=0$ ($f_p=0$). In this case, from the expression (36) of the main article, it follows that
	\begin{equation}
		\langle  \hat\varepsilon\rangle_{t}^Q =  \langle  \hat\varepsilon\rangle_{t}+\frac12  \alpha^2\Big(m^2\big( G_R^s(t,x) \big)^2 + \big(\partial_t G_R^s(t,x) \big)^2 + \big(\partial_x G_R^s(t,x) \big)^2 \Big).
	\end{equation}	
If the "smearing"\ function	has the form of (\ref{smeared-s}) than all necessary integrals can be done exactly. In particular, the  "smeared"\ retarded Green function is:
	\begin{multline}
		G_R^s(t,x) = \int dy\ \eta(y-x_q)G_R(t,x-y) =  
		-\theta(t) \int  \frac{dp}{2\pi} \frac{\sin(\omega_pt)}{\omega_p}
		e^{-ip( x-x_q)-\epsilon\omega_p}\\=\frac{i}{2\pi}\theta(t)\left(K_0\left(m\sqrt{(x-x_q)^2+(\epsilon-it)^2}\right)-K_0\left(m\sqrt{(x-x_q)^2+(\epsilon+it)^2}\right)\right).
	\end{multline}
The answer looks especially simple in the massless case $m\to 0$: 
	\begin{equation}
		G_R^s(t,x) =\frac{i}{4\pi}\log\left(\frac{(x-x_q)^2+(\epsilon+it)^2}{(x-x_q)^2+(\epsilon-it)^2}\right).
	\end{equation}
Than the energy density is:
	\begin{equation}
		\langle  \hat\varepsilon\rangle_{t}^Q =  \langle  \hat\varepsilon\rangle_{t}+\frac{\alpha^2}{4\pi^2}\left(\frac{\epsilon^2}{((x-x_q-t)^2+\epsilon^2)^2}+\frac{\epsilon^2}{((x-x_q+t)^2+\epsilon^2)^2}\right).\label{result-s}
	\end{equation}
	Now we can compare the above answer with the results of the conformal field theory. In order to do this, we need to turn to imaginary time $\tau=it$ and  introduce the complex coordinates $z=x+i\tau$, $\bar z=x-i\tau$. Quench is done by the operator $\hat Q(z,\bar z)$ with the conformal dimensions $(h,\bar h)$. In order to find the energy density, it is necessary to calculate the following ratio of the eucleadean  correlation functions \cite{AgeevHep-s,Caputa2-s}:
	\begin{equation}
		\delta\varepsilon(\tau,x)=-\frac{1}{2\pi}\frac{\langle  Q^\dag (x_q+i\epsilon,x_q-i\epsilon)(T(x+i\tau)+\bar T(x-i\tau)) Q (x_q-i\epsilon,x_q+i\epsilon)\rangle_E}{\langle  Q^\dag (x_q+i\epsilon,x_q-i\epsilon) Q (x_q-i\epsilon,x_q+i\epsilon)\rangle_E},
	\end{equation}
	and perform analytical continuation $\tau\to it$. Here $T(z)$ and $\bar T(\bar z)$ are holomorphic and antiholomorphic components of the stress-energy tensor, and a common multiplier is chosen to correspond to the energy density we calculated above. The correlation function in the numerator can be reduced to a pair correlation function using the conformal Ward identity \cite{Mussardo-s}, and the energy density become \cite{AgeevHep-s,Caputa2-s}:
	\begin{equation}
		\delta\varepsilon(\tau,x)=\frac{2h\epsilon^2}{\pi(x_q-x-i\epsilon-i\tau)^2(x_q-x+i\epsilon-i\tau)^2}+\frac{2\bar h\epsilon^2}{\pi(x_q-x+i\epsilon+i\tau)^2(x_q-x-i\epsilon+i\tau)^2}.
	\end{equation}
   After that, we need to perform an analytical continuation and use the fact that the conformal dimensions of the vertex operator
   $\hat {\mathcal V}_\alpha=:e^{i\alpha\hat \varphi}:$ equal to $(\alpha^2/(8\pi),\alpha^2/(8\pi))$ \cite{Mussardo-s}. As a result, we obtain exactly the answer of the expression (\ref{result-s}).

	\section{Quench with the operator $\hat \varphi(\bxq)$}
	
Quench with the operator $\hat Q(x)=\hat \varphi(x)$ has been considered in the work \cite{AgeevHep-s}. Although the expression (36) from the main article is not directly applicable to this case, the developed method can be easily generalised to treat such a quench. In this case, the equation (20) of the main article 
	\begin{equation}\label{qcsa-s}
		\langle \hat O\rangle_{t}^Q = 
		\int \mathfrak{D}\Phi(\bx)  \mathfrak{D} 
		\Pi(\bx) W[\Phi(\bx),\Pi(\bx)]\Qsm O[\phi_c(t,\Phi(\bx),\Pi(\bx))]
	\end{equation}	
    is still correct if 
	\begin{equation}
		\Qsm = \mathcal N^2\left(\Phi_s+i\frac{\hbar}{2}\frac{\delta}{\delta\Pi_s}\right)\left(\Phi_s-i\frac{\hbar}{2}\frac{\delta}{\delta\Pi_s}\right)=\mathcal N^2\left(\Phi_s^2+\frac{\hbar^2}{4}\frac{\delta^2}{\delta\Pi_s^2}\right).
	\end{equation}
Here normalisation constant $\mathcal N$ is defined from the condition $\langle 1\rangle_{t}^Q =1$: 
	\begin{equation}
		\mathcal N^{-2} = 
		\int \mathfrak{D}\Phi(\bx)  \mathfrak{D} 
		\Pi(\bx) W[\Phi(\bx),\Pi(\bx)]\Phi_s^2=\langle \Phi_s^2\rangle_{i.c.}.
	\end{equation}	
With the above condition, further calculations are performed analogously to those in the main article. The energy density after quench equals:
	\begin{multline}
		\langle  \hat\varepsilon\rangle_{t}^Q = \frac{1}{2\langle \Phi_s^2\rangle_{i.c.}}\big\langle \Phi_s^2(m^2\phi_c^2(t,x)+(\partial_t\phi_c(t,x))^2+(\partial_x\phi_c(t,x))^2)\big\rangle_{i.c.}\\+\frac{\hbar^2}{4\langle \Phi_s^2\rangle_{i.c.}}\Big(m^2\big( G_R^s(t,x) \big)^2
		+ \big(\partial_t G_R^s(t,x) \big)^2 + \big(\partial_x G_R^s(t,x) \big)^2 \Big).
	\end{multline}
 If we use vacuum initial conditions, then the Wick theorem is applicable to the calculation of the averages. The result is:
	\begin{multline}
		\langle  \hat\varepsilon\rangle_{t}^Q =\langle  \hat\varepsilon\rangle_{t}^Q - \frac{1}{\langle \Phi_s^2\rangle_{i.c.}}\Big(m^2\big( G_K^s(t,x) \big)^2
		+ \big(\partial_t G_K^s(t,x) \big)^2 + \big(\partial_x G_K^s(t,x) \big)^2 \Big)\\+\frac{\hbar^2}{4\langle \Phi_s^2\rangle_{i.c.}}\Big(m^2\big( G_R^s(t,x) \big)^2
		+ \big(\partial_t G_R^s(t,x) \big)^2 + \big(\partial_x G_R^s(t,x) \big)^2 \Big).
	\end{multline}
	As in the previous section, for the vacuum initial state the calculations can be done analytically. In the case of regularization used in the work \cite{AgeevHep-s}, the "smearing"\ function should be chosen in the form (\ref{smeared-s}). So, the energy density is:
	\begin{equation}
		\langle \Phi_s^2\rangle_{i.c.}=i\int dy dz\ \eta(y-x_q)\eta(z-x_q)G_K^s(0,y-z)=\hbar\int\frac{dp}{2\pi}\frac{1}{2\omega_p}e^{-2\epsilon\omega_p}=\frac{\hbar }{2\pi}K_0(2m\epsilon),
	\end{equation}
	\begin{equation}
		iG_K^s(t,x)+i\frac{\hbar}{2}G_R^s(t,x)=\hbar\int\frac{dp}{2\pi}\frac{1}{2\omega_p}e^{-ip(x-x_q)-it\omega_p-\epsilon\omega_p}=\frac{\hbar }{2\pi}K_0\left(m\sqrt{(x-x_q)^2+(\epsilon+it)^2}\right),
	\end{equation}
	\begin{multline}
		\langle  \hat\varepsilon\rangle_{t}^Q =\langle  \hat\varepsilon\rangle_{t}^Q+\frac{m^2\hbar}{2\pi K_0(2m\epsilon)}\left|K_0\left(m\sqrt{(x-x_q)^2+(\epsilon+it)^2}\right)\right|^2\\+\frac{m^2\hbar\left((x-x_q)^2+t^2+\epsilon^2\right)}{2\pi K_0(2m\epsilon)}\left|\frac{K_1\left(m\sqrt{(x-x_q)^2+(\epsilon+it)^2}\right)}{\sqrt{(x-x_q)^2+(\epsilon+it)^2}}\right|^2.
	\end{multline}
   Exactly the same answer was obtained in the work \cite{AgeevHep-s} with the help of eucleadean time calculations and analytical continuation. Therefore, the results obtained for the vacuum initial state can be calculated using the method of the main article with the correct choice of the "smearing"\ function $\eta(x)$. It is important to note that the presented method is more general and can be applied to different quench operators and arbitrary initial conditions.

\end{document}